\documentclass[prl,aps,amssymb,showpacs,twocolumn]{revtex4}

\usepackage{dcolumn}
\usepackage{graphicx}
\usepackage{bm}

\usepackage{amsmath}
\usepackage{amssymb}
\usepackage{amsthm}
\usepackage{amsfonts}
\usepackage{enumerate}
\usepackage{latexsym}

\input{epsf}

\begin{document}
\title{A minimal two-band model for the superconducting Fe-pnictides}
\author{S. Raghu$^1$, Xiao-Liang Qi$^1$, Chao-Xing Liu$^{2,1}$, D.J. Scalapino$^3$ and Shou-Cheng Zhang$^1$}
\affiliation{$^1$Department of Physics, McCullough Building,
Stanford University, Stanford, CA 94305-4045}
 \affiliation{$^2$Center for Advanced
Study, Tsinghua University, Beijing, 100084, R. P. China}
\affiliation{$^3$Department of Physics, University of California,
Santa Barbara, CA 93106-9530}
\date{\today}
\begin{abstract}
Following the discovery of the Fe-pnictide superconductors, LDA band structure
calculations showed that the dominant contributions to the spectral weight near the
Fermi energy came from the Fe 3d orbitals.  The Fermi surface is
characterized by two hole surfaces around the $\Gamma$ point
and two electron surfaces around the M point of the 2 Fe/cell
Brillouin zone.  Here, we describe a 2-band model
that reproduces the topology of the LDA Fermi surface and exhibits both
ferromagnetic and $q=(\pi,0)$ spin density wave (SDW) fluctuations.
We argue that this minimal model contains the essential low energy
physics of these materials.
\end{abstract}


\maketitle \emph{Introduction - } The recent discovery of
superconductivity in a family of Fe-based oxypnictides with large
transition temperatures \cite{Kamihara2008, ren2008, Chen2008,
Chen2008A,Chen2008B,wen2008} has led to tremendous activity aimed at
identifying the mechanism for superconductivity in these materials.
Preliminary experimental results including specific heat
\cite{Mu2008}, point-contact spectroscopy \cite{Shan2008} and
high-field resistivity \cite{Hunte2008, Zhu2008} measurements
suggest the existence of unconventional superconductivity in these
materials.  Furthermore, transport\cite{dong2008} and neutron
scattering\cite{delacruz2008} measurements have shown the evidence
of magnetic order below $T=150K$. An experimental determination of
the orbital and spin state of the Cooper pairs, however, has not yet
been made.

The high transition temperatures and the electronic structure of the
Fe-pnictide superconductors suggest that the pairing interaction is
of electronic origin.\cite{Boeri2008} First-principle band structure
calculations \cite{Singh2008,Xu2008,Mazin2008,Haule2008} have shown
that superconductivity in these materials is associated with the
Fe-pnictide layer, and that the density of states (DOS) near the
Fermi level gets its maximum contribution from the Fe-3d orbitals.
The consensus based on these calculations is that the Fermi surface
consists of two hole pockets and two electron pockets. Calculations
from Ref. \cite{Xu2008} also show van Hove singularities which might
be responsible for enhanced ferromagnetic fluctuations. The bare
magnetic spin susceptibility determined from these bands exhibits
both ferromagnetic $q \sim 0$ and finite q SDW peaks.

Several tight binding models for the band structure have been
proposed.  Cao et al. \cite{Cao2008} used 16 localized Wannier
functions to construct a tight-binding effective Hamiltonian. Kuroki
et al. \cite{Kuroki2008} have used a 5 orbital tight binding model
to fit the band structure near the Fermi energy. Others have
introduced generic 2-band models \cite{Dai2008,Han2008,Li2008}.
However, the relationship of these latter models to the multiple
Fermi surface electron and hole pockets found in LDA calculations is
unclear.  Since it appears likely that these multiple Fermi surfaces
play an essential role in determining the momentum dependence of the
spin and orbital fluctuations which would mediate an electronic
pairing mechanism, we would like to construct a minimal model that
exhibits a Fermi surface similar to that obtained from band
structure calculations.  

This model has two
orbitals per site on a two dimensional square lattice.
By adjusting the one-electron hopping parameters and the chemical
potential one can obtain a Fermi surface which has the same
topology as found from the band structure calculations.  The
non-interacting spin susceptibility also exhibits both
ferromagnetic and finite $q$ SDW peaks.  With the addition of an
onsite intra-orbital and inter-orbital Coulomb interactions, and
an intra-orbital Hunds rule coupling, this model represents what we
believe is a minimal model for describing the low energy physics
of these materials. 
In addition, the relative simplicity of
this model should be useful in the phenomenological analysis of
experiments related to the gap symmetry \cite{Graser2008}
 and in
numerical density-matrix renormalization group (DMRG) and dynamic cluster studies.

\textit{Model Hamiltonian - } The structure of the FeAs layer of
LaOFeAs viewed along the c-axis is illustrated in Fig.~1a.
\begin{figure}[ht]
\includegraphics [width=8cm,clip,angle=0]{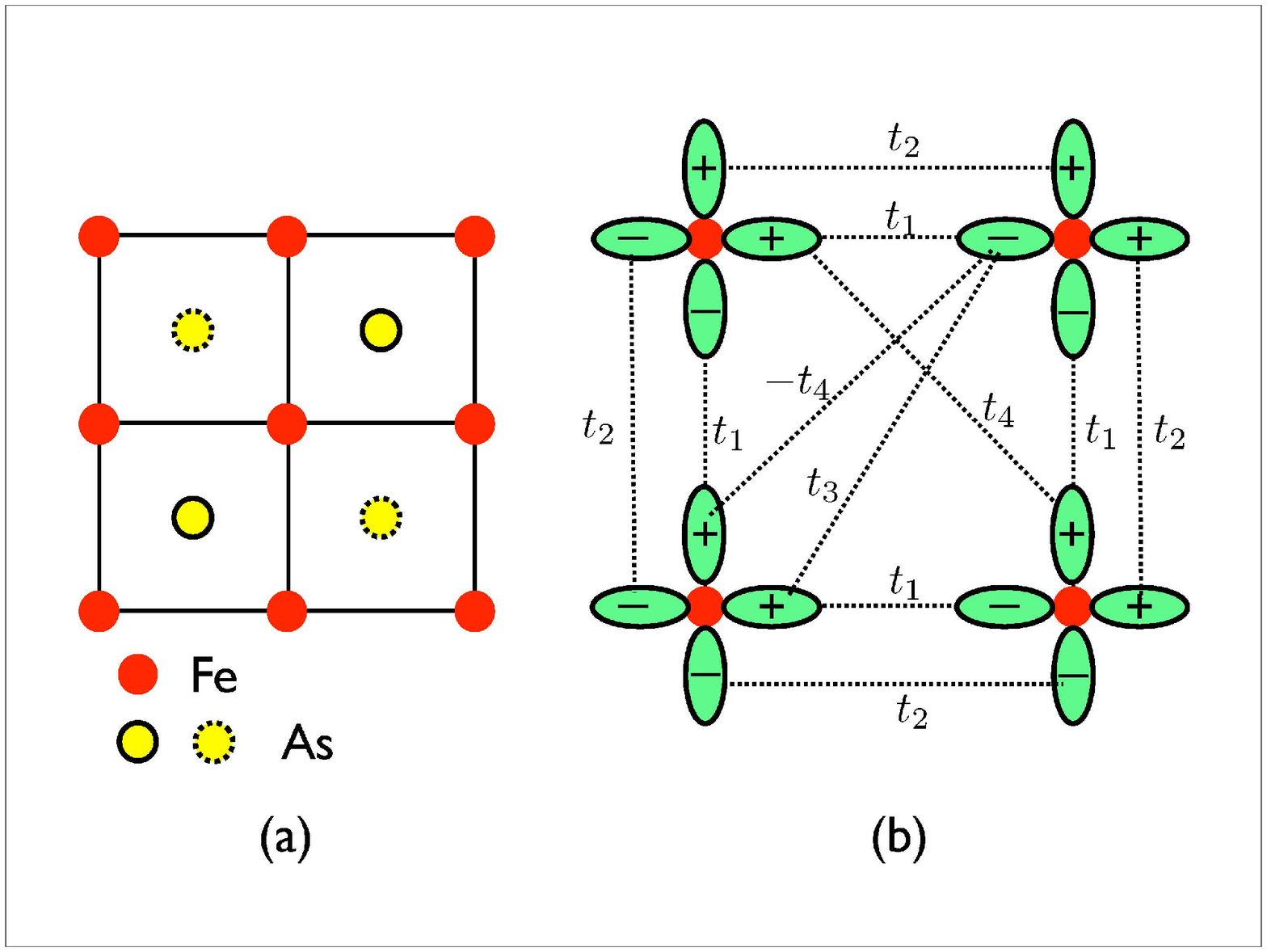}
\caption{(a) The Fe ions form a square lattice and the
crystallographic unit cell contains two Fe and two As ions.  The As
ions are located either directly above (solid circles) or below
(dashed circles) the faces of the Fe square array. (b)A schematic
showing the hopping parameters of the two-orbital $d_{xz}-d_{yz}$
model on a square lattice. Here $t_1$ is a near neighbor hopping
between $\sigma$-orbitals and $t_2$ is a near neighbor hopping
between $\pi$-orbitals. We also include a second-neighbor hopping
$t_4$ between different orbitals and a second-neighbor
hopping $t_3$ between similar orbits.} \label{tb}
\end{figure}
The Fe ions form a square lattice which is interlaced with a second
square lattice of As ions. These As ions sit in the center of each
square plaquette of the Fe lattice  and are displaced above and
below the plane of the Fe ions as indicated in the figure. This
leads to two distinct Fe sites and a crystallographic unit cell
which contains two Fe and two As ions. As shown by various band
structure calculations, the main contribution to the density of
states within several eV of the Fermi surface comes from the Fe 3d
states which disperse only weakly in the $z$-direction.  The
 3d Fe orbitals hybridize among themselves and
through the As $p$ orbitals leading to a complex of bands.
However, as noted in Ref. \cite{Mazin2008}, the bandstructure near
the Fermi level is relatively simple in the unfolded 1Fe/cell BZ
where it primarily involves three Fe orbitals $d_{xz}$, $d_{yz}$ and
$d_{xy}$ (or $d_{x^2-y^2}$). Based upon this observation and by
making the further approximation that the role of the $d_{xy}$
($d_{x^2-y^2}$) orbit can be replaced by a next near neighbor
hybridization between $d_{xz},d_{zy}$ orbitals, we consider a
two-dimensional square lattice with two degenerate
``$d_{xz},d_{yz}$" orbitals per site. While one may well need a
third orbit to control the relative sizes and eccentricities of the
electron and hole pockets, we find that a two-orbital model can lead
to a Fermi surface  which resembles that obtained in the
bandstructure calculations.

The tight-binding parameters of the 2-orbital model that we will
study are illustrated in Fig. \ref{tb}.
\begin{figure}[ht]
\includegraphics [width=9cm,clip,angle=0]{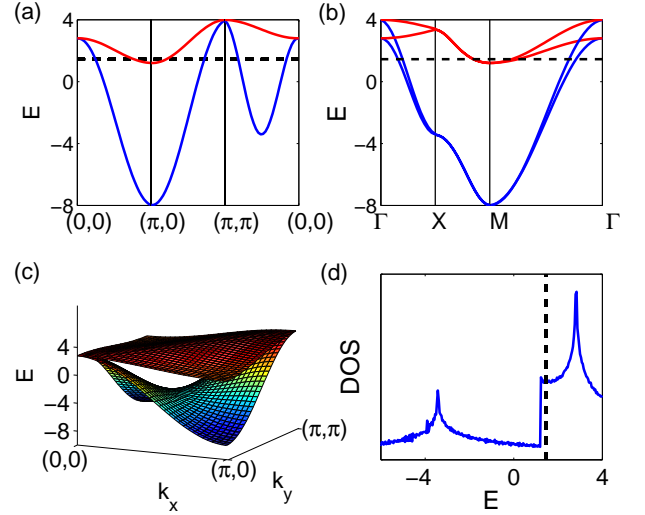}
\caption{(a) The band structure of the two-band model with 
$t_1=-1,t_2 = 1.3, t_3=t_4=-0.85$ and $\mu = 1.45$, 
plotted along the path
$(0,0)\rightarrow (\pi,0)\rightarrow (\pi,\pi)\rightarrow (0,0)$ as
shown in Fig. \ref{fs} (a) by the black dashed lines. (b) The band
structure folded to the small BZ, with the $\Gamma,X,M$ defined in
the small BZ as shown in Fig. \ref{fs} (b). (c) The two-d band
structure for $k_x,k_y\in[0,\pi]$. A saddle point exists for each
band. (d) The density of states of the two band model, with two Van
Hove singularities. The dashed line shows the fermi level
corresponding to our choice of $\mu=1.45$.} \label{band}
\end{figure}
\begin{figure}[ht]
\includegraphics [width=5.5cm,clip,angle=0]{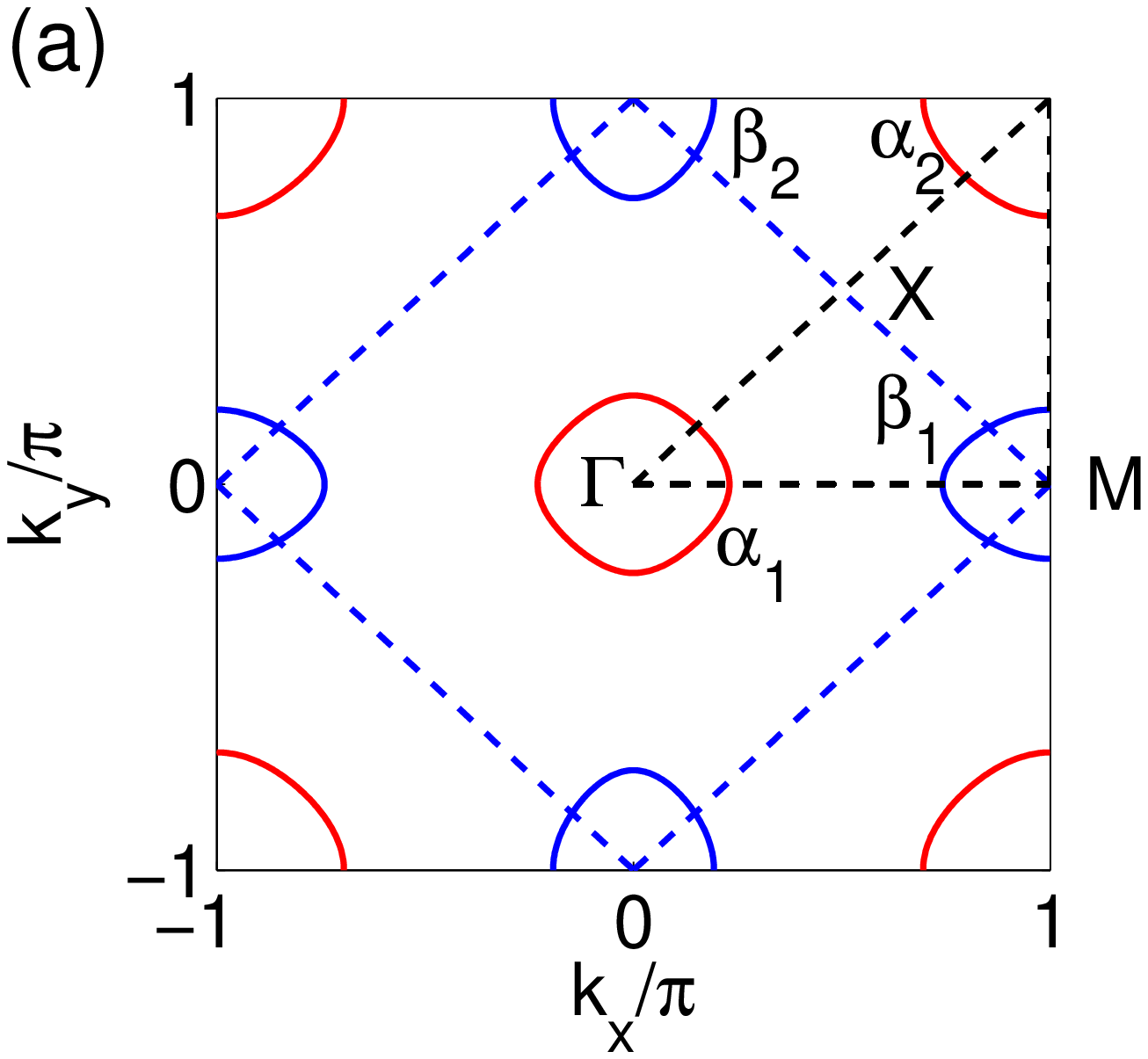}\\
\includegraphics [width=5.5cm,clip,angle=0]{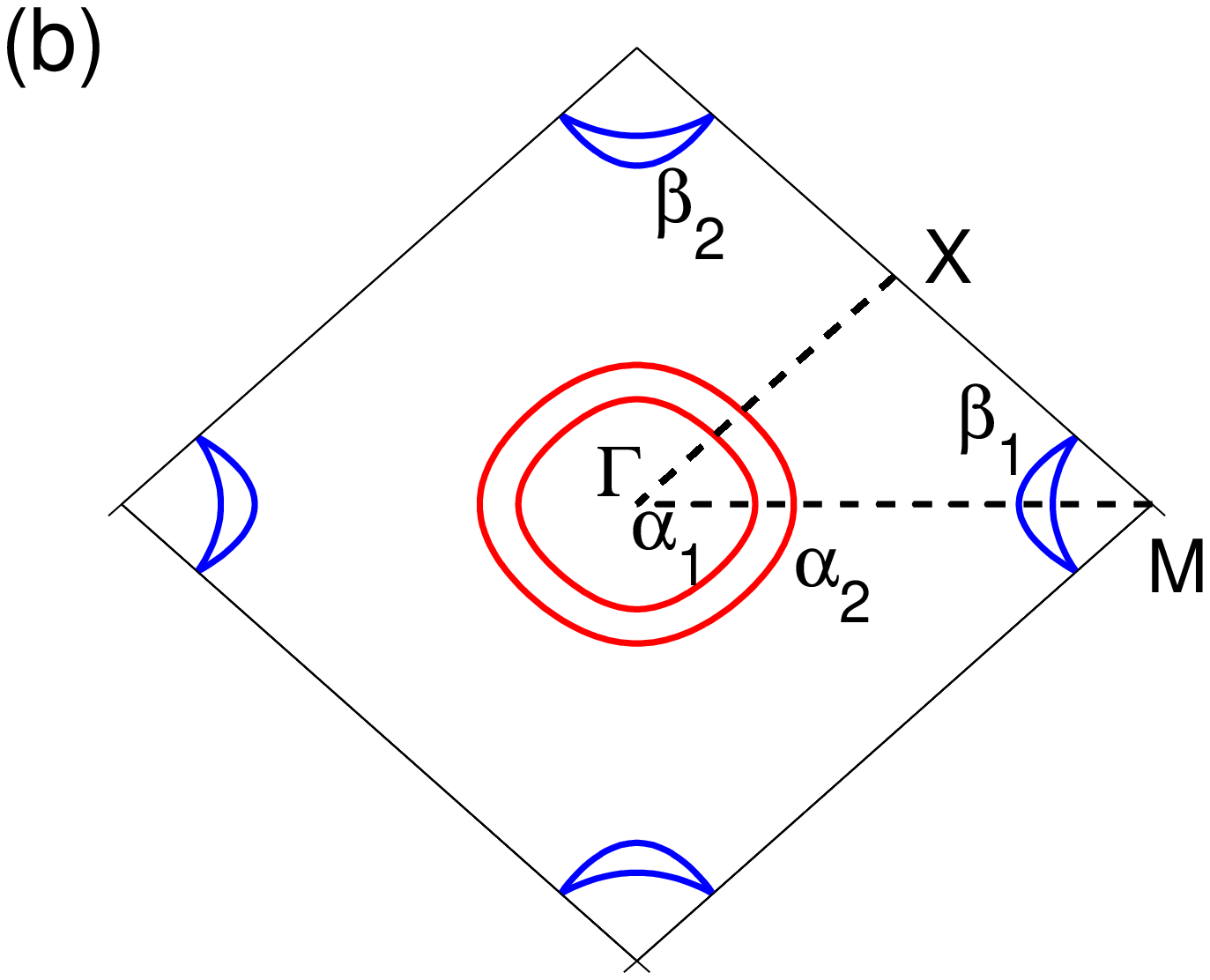}
\caption{(a) The Fermi surface of the 2-orbital model on the large
1Fe/cell BZ. Here, $\alpha_{1,2}$ surfaces are hole Fermi pockets
given by $E_-(\bm k_f) = 0$ and $\beta_{1,2}$ are electron Fermi
pockets by $E_+(\bm k_f) = 0$. The dashed square indicates the BZ of
the 2Fe/cell. (b) The Fermi surface folded down into the 2
Fe/cell BZ consists of two $\alpha$ surfaces around $\Gamma$ and two
elliptically deformed $\beta$ surfaces around the $M$ point. Here
the parameters are the same as in Fig. \ref{band}}
\label{fs}
\end{figure}
It is convenient to introduce a two-component field
\begin{equation}
\psi_{ks}={d_{xs}(k)\choose d_{ys}(k)} \label{eq:1}
\end{equation}
Here $d_{xs}(k)$ ($d_{ys}(k)$) destroys a $d_{xz}$ ($d_{yz}$)
electron with spin $s$ and wave vector $k$. Then the tight binding
part of the Hamiltonian can be written as
\begin{equation}
H_0=\sum_{ks}\psi^+_{ks}\left[\left(\varepsilon_+(k)-\mu\right)1+\varepsilon_-(k)
\tau_3+\varepsilon_{xy}(k)\tau_1\right]\psi_{ks}, \label{Htb}
\end{equation}
with $\tau_i$ the Pauli matrices and
\begin{eqnarray}
  \varepsilon_\pm(k) & = & \frac{\varepsilon_x(k)\pm\varepsilon_y(k)}{2}, \nonumber\\
    \varepsilon_x(k) & = & -2t_1\cos k_x-2t_2\cos k_y-4t_3\cos k_x\cos k_y,\nonumber\\
    \varepsilon_y(k) & = & -2t_2\cos k_x-2t_1\cos k_y-4t_3\cos k_x\cos k_y,\nonumber\\
    \varepsilon_{xy}(k) & = & -4t_4\sin k_x\sin k_y.\nonumber
\end{eqnarray}
\label{eq:3}
The one electron Matsubara Green's function is given by
\begin{equation}
\hat{G}_s (\bm k, i \omega_n ) = \frac{\left( i \omega_n - \epsilon_+(\bm k) \right) \hat{1} - \epsilon_-(\bm k) \hat{\tau}_3 - \epsilon_{xy} (\bm k) \hat{\tau}_1}{\left(i \omega_n - E_+(\bm k) \right) \left(i \omega_n - E_-(\bm k) \right)}
\end{equation}
with
\begin{equation}
E_{\pm}(\bm k) = \epsilon_+(\bm k) \pm \sqrt{\epsilon_-^2(\bm k) +
\epsilon_{xy}^2(\bm k)} - \mu\label{Ek}
\end{equation}

In Figure \ref{band} we show the band structure of the model for a
specific choice of hopping parameters $t_1=-1,t_2 = 1.3,
t_3=t_4=-0.85$, in units of $\vert t_1 \vert$.
The folded energy spectrum in Fig.
\ref{band} (b) shows the band structure in the 2 Fe/cell zone.  
Due to the saddle points in the energy spectrum (as
shown in Fig. \ref{band} (c)), there are two Van Hove singularities
in the density of states, which also qualitatively agrees with the
LDA results.\cite{Xu2008} In Figure \ref{fs} we show the Fermi
surface for the same set of parameters. On the large BZ (Fig.
\ref{fs}a) associated with our model which has 1 Fe/unit cell, there
are two hole Fermi pockets labeled $\alpha_1$ and $\alpha_2$ defined by
$E_-(\bm k) = 0$, and two electron Fermi pockets $\beta_1$ and $\beta_2$
defined by $E_+(\bm k) = 0$. To compare with band structure
calculations, one must fold the large BZ into a smaller one which is
dual to the crystallographic unit cell containing two Fe atoms. The
dashed square in Fig. \ref{fs} (a) marks this smaller zone and in
Fig. \ref{fs} (b) we show what happens as the $\alpha_{1,2}$ and
$\beta_{1,2}$ bands of Fig. \ref{fs} (a) are folded back into the
2Fe/cell BZ.  One sees that this gives Fermi surfaces with
 the same topology that is obtained from LDA 
 band structure calculations \cite{footnote1}.

\emph{One-loop spin susceptibility.--} Now we study the one-loop
spin-susceptibility for the tight-binding model (\ref{Htb}). Due to
the existence of two degenerate orbitals in our model, the spin
susceptibility also has orbital indices, and is defined by
\begin{eqnarray}
\chi_{st}({\bf q},i\Omega)=\int_0^\beta d\tau e^{i\Omega
\tau}\left\langle T_\tau {\bf S}_s({\bf -q},\tau)\cdot {\bf
S}_t({\bf q},0)\right\rangle\label{chi0}
\end{eqnarray}
here $s,t=1,2$ label the orbital indices, and ${\bf S}_s({\bf
q})=\frac12\sum_{\bf k}\psi_{s\alpha}^\dagger({\bf
k+q})\vec{\sigma}_{\alpha\beta}\psi_{s\beta}({\bf k})$ is the spin
operator for the orbital labeled by $s$. The physical spin susceptibility is
given by $$\chi_{\rm S}({\bf q},i\Omega)=\sum_{s,t}\chi_{st}({\bf
q},i\Omega).$$ The one loop contribution to the spin susceptibility
can be obtained as
\begin{eqnarray}
\chi_{\rm S}({\bf q},i\Omega)&=&- \frac{T}{2N} \sum_{\bm k,
\omega_n} {\rm Tr} \left[ G(\bm k + \bm q, i \omega_n + i \Omega)
G(\bm k , i
\omega_n) \right]\nonumber\\
&=&-\frac{1}{2N}\sum_{{\bf k},\nu,\nu'}\frac{\left|\left\langle {\bf
k+q},\nu|{{\bf k},\nu'}\right\rangle\right|^2}
{i\Omega+E_{\nu,{\bf k+q}}-E_{\nu',{\bf k}}}\nonumber\\
& &\cdot\left(n_F(E_{\nu,{\bf k+q}})-n_F(E_{\nu',{\bf k}})\right)
\end{eqnarray}
Here $E_{\nu{\bf k}},~\nu=+1 (-1)$ is the eigenvalue of the upper
(lower) band given by Eq. (\ref{Ek}), and $\left|{\bf
k},\nu\right\rangle$ the corresponding eigenvector.
$n_F(E)=1/(e^{\beta E}+1)$ is the fermi distribution function.

\begin{figure}[ht]
\includegraphics [width=4.5cm,clip,angle=0]{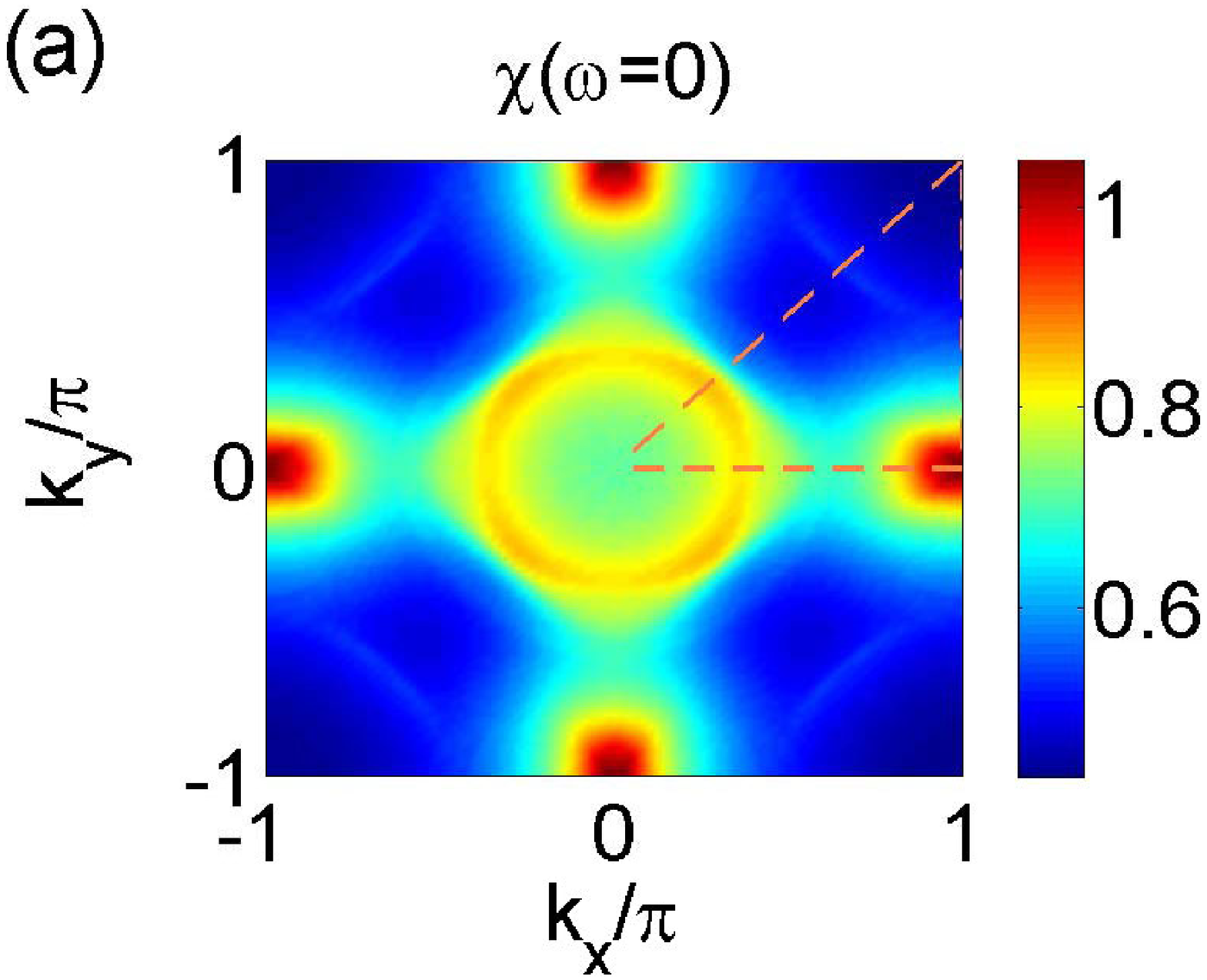}\includegraphics [width=4.5 cm,clip,angle=0]{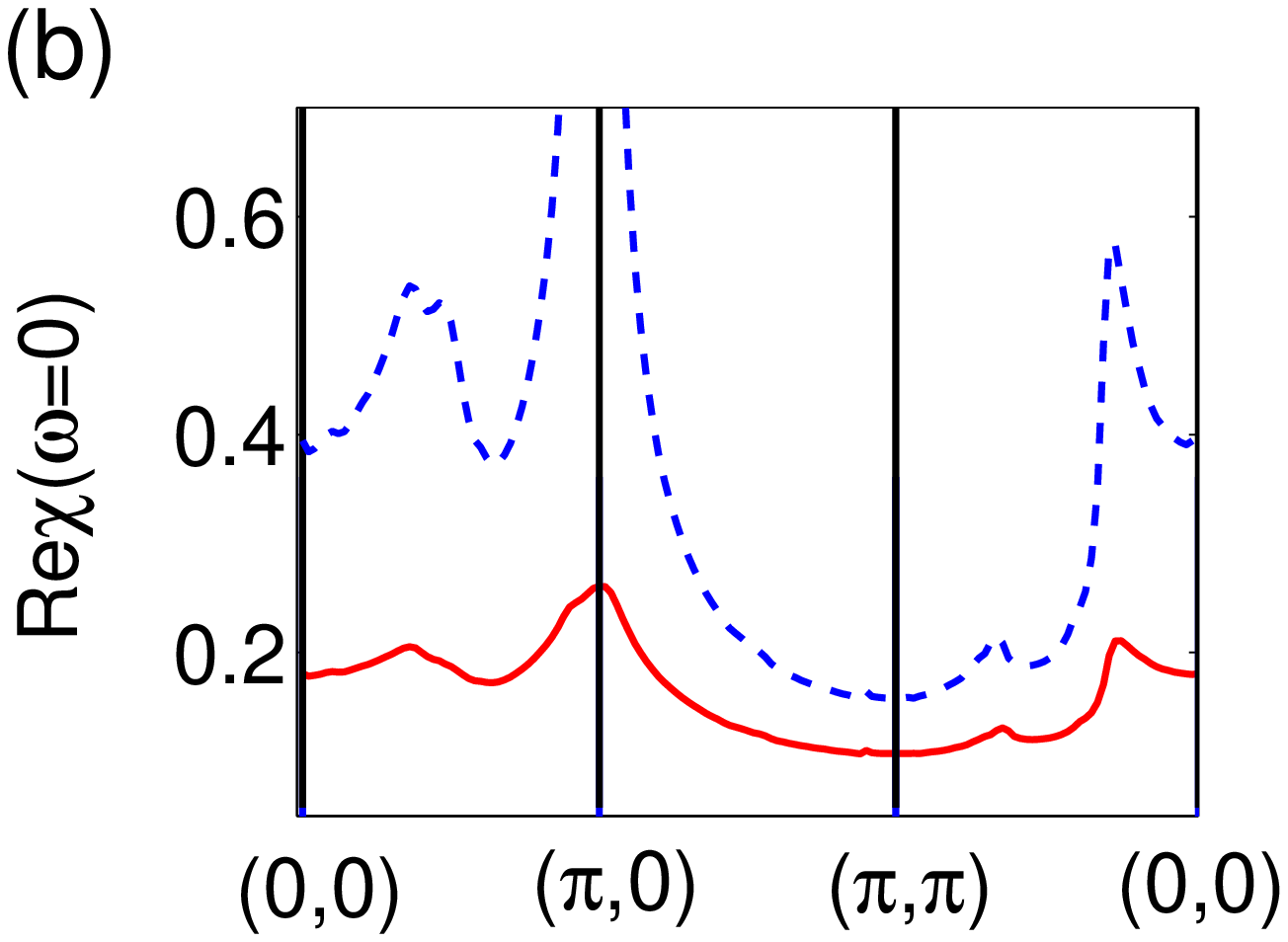}
\caption{(a) The $\omega=0$ bare spin susceptibility $\chi_{\rm
S}({\bf q})$ versus ${\bf q}$ for the same tight-binding parameters
as used in Fig.~\protect\ref{fs}. (b) The bare spin susceptibility
$\chi_{\rm S}({\bf q})$ (red solid line) and the RPA spin
susceptibility $\chi_{\rm S}^{\rm RPA}({\bf q})$ for
$U=V=6,~J=0$ (blue dashed line) along the $(0,0)\rightarrow
(\pi,0)\rightarrow (\pi,\pi)\rightarrow (0,0)$ path in BZ, as shown
by the dashed line in subfigure (a).} \label{fig:3}
\end{figure}

\begin{figure}[ht]
\includegraphics [width=4.5cm,clip,angle=0]{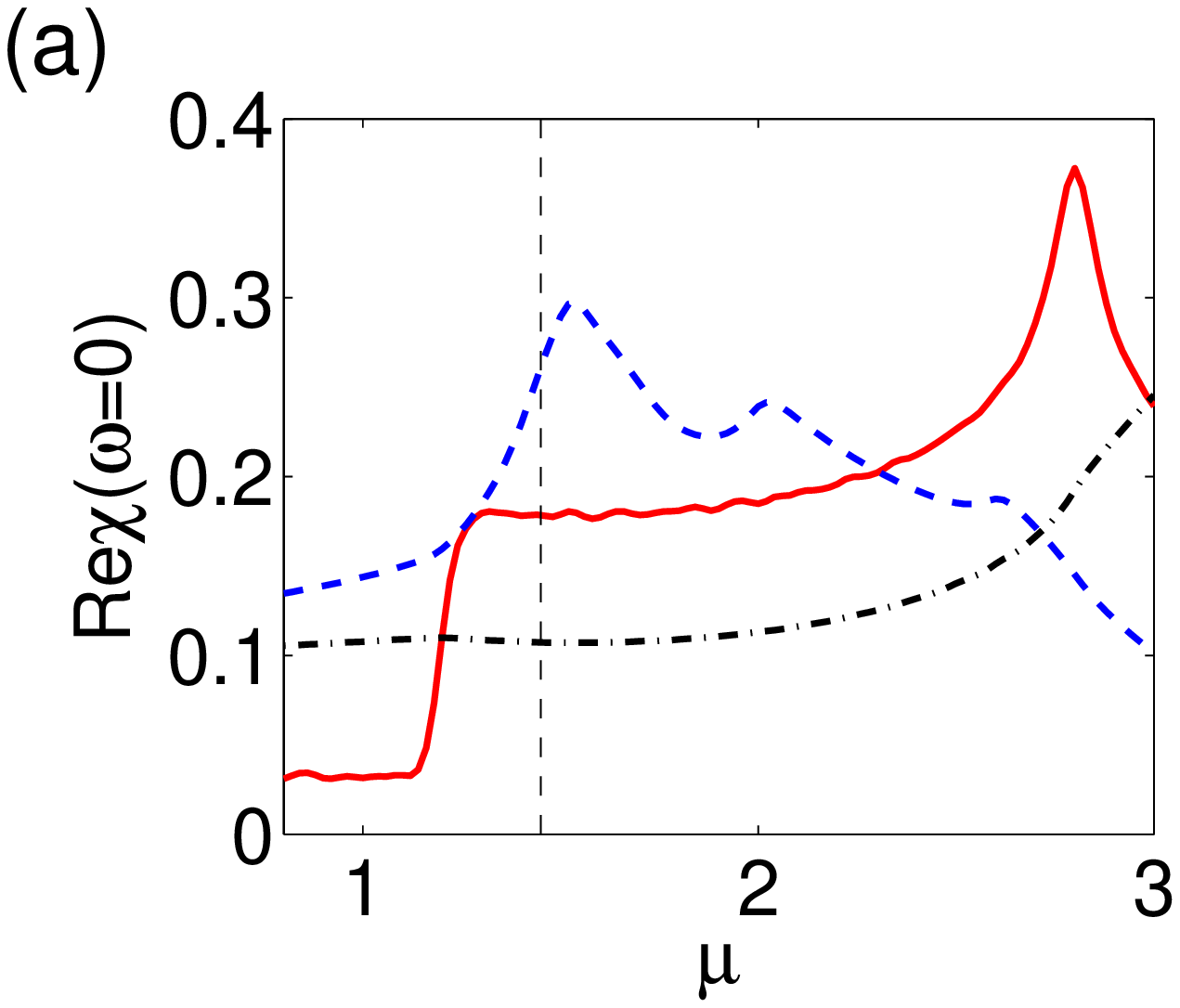}\includegraphics [width=4.5cm,clip,angle=0]{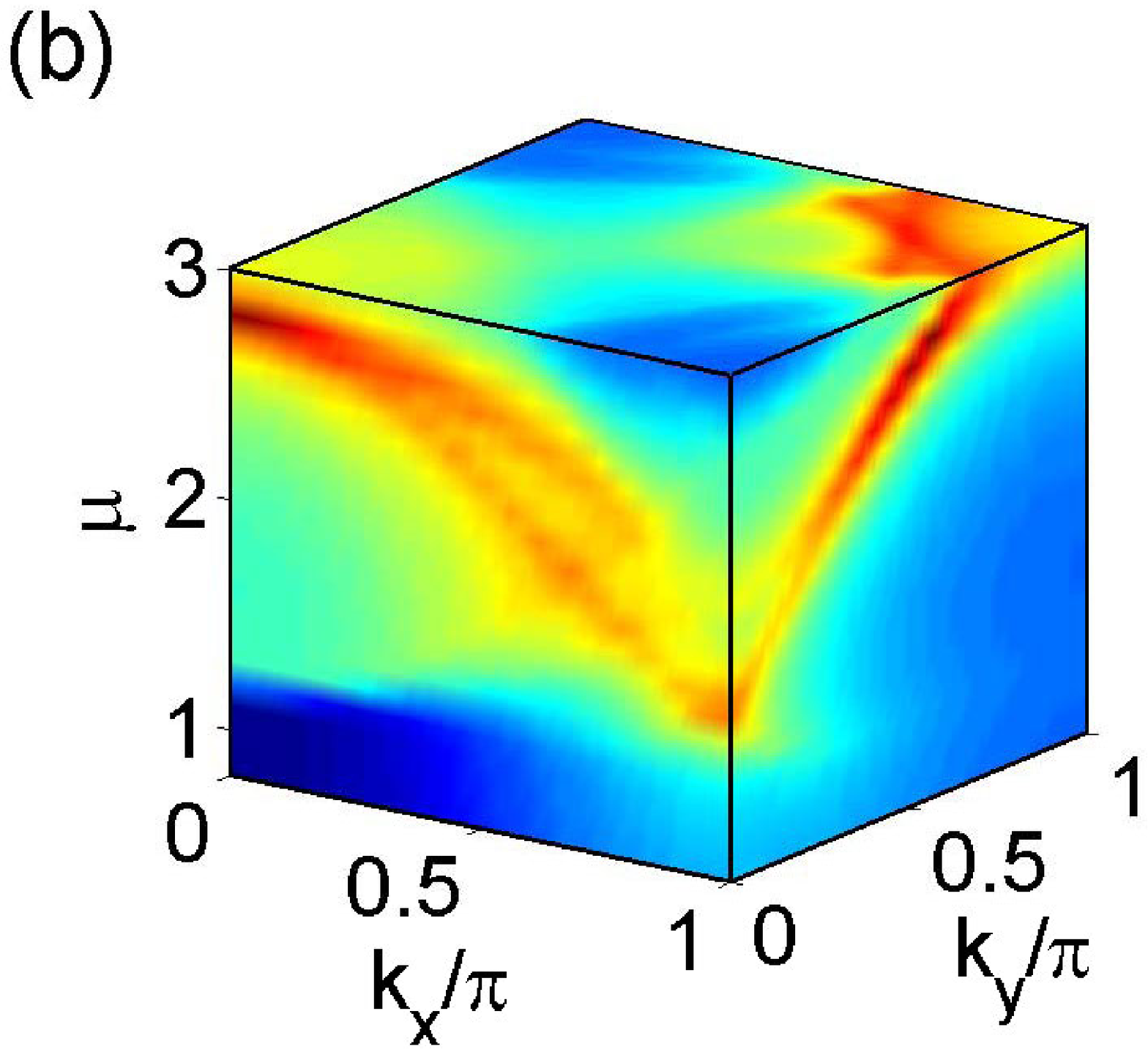}
\caption{(a) The chemical potential dependence of ${\rm Re}\chi({\bf
q},\omega=0)$ with ${\bf q}=(0,0)$, $(0,\pi)$ and $(\pi,\pi)$. The
vertical dashed line shows the chemical potential $\mu=1.45$ which
we are working on. (b) The three-d color plot of spin susceptibility
as a function of $k_x,k_y$ and chemical potential $\mu$. The hot
points in BZ shifts from $(\pi,0)$ to $(0,0)$ and $(\pi,\pi)$ 
with electron doping as $\mu$ increases.  
The deepest red region shows the Van Hove
singularities.} \label{chiVSmu}
\end{figure}

Fig.~\ref{fig:3} shows a plot of the static spin susceptibility
$\chi_S(q,0)$ versus $q$, where one can see the structure associated
with the various nesting points and density of states features. For
our choice of parameters, the largest value of $\chi_0(q)$ occurs
around $q=(\pi,0)$ and $(0,\pi)$, which suggests a transition to an
antiferromagnetic (AFM) ordered phase at some critical interaction
strength. This is also in agreement with the result of band
structure calculations\cite{Mazin2008,zhang2008}. 
A recent neutron scattering experiment has confirmed that such
a peak develops below $T \sim 150K$ \cite{delacruz2008}. Such a peak
in the spin susceptibility comes from the nesting between the
electron and hole Fermi pockets, which can be seen from the chemical
potential dependence of the spin susceptibility. As shown in Fig.
\ref{chiVSmu}, the spin susceptibility at $q=(0,0)$ jumps
discontinuously at $\mu\sim 1.2$, which follows the behavior of the
density of states shown in Fig. \ref{band} (d), and corresponds to
the onset of electron Fermi pockets. At the same time, the
$q=(0,\pi)$ spin susceptibility is also enhanced significantly due
to the nesting effect. When the chemical potential is increased
further, the fermi level gets closer to the Van Hove singularity,
and the hot point of the spin susceptibility is shifted gradually to the
neighborhood of $(0,0)$ and $(\pi,\pi)$, as shown in Fig.
\ref{chiVSmu} (b).

{\em The RPA spin susceptibility.}--- Now we consider
the effect of electron-electron interaction in this model. For the
two $d$ orbitals we considered, the generic form of the on-site
interaction can be written as
\begin{eqnarray}
H_{\rm
int}=\sum_i\left(U\sum_sn_{is\uparrow}n_{is\downarrow}+Vn_{i1}n_{i2}-J{\bf
S}_{i1}\cdot {\bf S}_{i2}\right)
\end{eqnarray}
with $U$ and $V$ the intraband and interband Coulomb repulsion,
and $J$ the Hunds rule coupling. For an isolated ${\rm Fe}$ atom,
 the intraband $U$ and interband $V$ are similar in magnitude,
  and $J$ is an order of magnitude smaller
\cite{Schnell2003}. Thus we expect the $U$ and $V$ to be the
dominant terms in the interaction. We suggest that $H=H_{\rm 0}+ H_{\rm
int}$ represents a minimal model for the Fe-pnictides superconductors.

Next we will study the effect of such interactions on the spin fluctuations within 
RPA.  Due to the two
band nature of the model we considered, the RPA correction should be
calculated for the generic spin susceptibility $\chi_{st}({\bf
q},i\Omega)$ defined in Eq. (\ref{chi0}), which is determined by the
following matrix equation:
\begin{eqnarray}
\chi^{\rm RPA}({\bf q},i\Omega)=\chi_0({\bf
q},i\Omega)\left(\mathbb{I}-\Gamma \chi_0({\bf
q},i\Omega)\right)^{-1}\label{chiRPA}
\end{eqnarray}
Here $\chi_0$ is the $2\times 2$ matrix formed by the intra-orbital
and inter-orbital spin susceptibility defined in Eq. (\ref{chi0}),
and $\Gamma$ is the interaction vertex defined by
\begin{eqnarray}
\Gamma=\left(\begin{array}{cc}U&J/2\\J/2&U\end{array}\right)\label{vertex}
\end{eqnarray}
We note that the interband interaction $V$ does not
contribute to the RPA response when only the spin fluctuations are
considered.

In the following we set $J=0$, which makes
the interaction vertex $\Gamma$ in Eq. (\ref{vertex}) proportional
to the identity. For the tight-binding model parameters used in Fig.
\ref{fs} and $U=6$, we obtain the RPA spin susceptibility shown
in Fig. \ref{fig:3} (b) by the dashed line. As expected, the spin
susceptibility is enhanced around the hot points
$(\pi,0)$ and $(0,\pi)$. We have also carried out the RPA calculation for a 
finite Hunds rule coupling $J>0$, and find that the spin
fluctuations are enhanced by increasing $J$, but the structure of
$\chi({\bf q})$ remains qualitatively the same.

In conclusion, we have described a minimal model for the
Fe-pnictides which we believe contains the essential low energy
physics of these materials. This model consists of a two dimensional
square lattice of sites with each site having two degenerate
orbitals. By fitting the tight binding parameters, one can obtain a
band structure which, after folded to the 2Fe/cell BZ, exhibits two
hole pockets around the $\Gamma$ point and two electron pockets
around the $M$ point. The electron-electron interactions are taken
to be onsite intra-orbital and inter-orbital Coulomb interactions
$U$ and $V$ and an onsite Hund's rule coupling $J$. The structure of
the bare spin susceptibility is peaked around $(\pi,0)$ for the
parameters we chose to fit the fermi surface. Such AFM spin
fluctuations also leads to the possibility of non-conventional
superconductivity, which we will discuss in a separate work.
Different types of spin or orbial orders and superconductivity can
possibly occur for different fillings. Therefore, we conclude that
this model contains a rich variety of magnetic, orbital and pairing
correlations.

We would like to acknowledge X. Dai, Z. Fang and H. J. Zhang for
many insightful discussions and for generous sharing of their
unpublished work. We acknowledge helpful discussions with S.
Kivelson, R. Martin, I. Mazin, T. Schulthess, D. Singh and H. Yao.
We would also like to thank the authors of Ref. \cite{Graser2008}
for sending us their paper prior to submission. This work is
supported by the NSF under grant numbers DMR-0342832, the US
Department of Energy, Office of Basic Energy Sciences under contract
DE-AC03-76SF00515, the center for nanophase material science, ORNL
(DJS) and the Stanford Institute for Theoretical Physics (SR, DJS).

\bibliography{feasV2}

\begin{thebibliography}{26}
\bibitem[{foo()}]{footnote1}
\bibinfo{note}{The relative size of the Fermi velocity associated with the
  $\alpha$ and $\beta$ bands is not accurate and additional orbitals or a
  phenomenological adjustment would be required to fit the band structure
  precisely.}
\end{thebibliography}

\end{document}